\documentclass[12pt]{iopart}

\usepackage{booktabs}
\usepackage{graphicx}
\usepackage{utfsym}
\usepackage{url}
\usepackage{esvect}
\begin{document}

\title[]{Climate change impact on photovoltaic power potential in South America}

\author{Gabriel Narvaez\textsuperscript{1}, Michael Bressan\textsuperscript{1}, Andres Pantoja\textsuperscript{2}, Luis Felipe Giraldo\textsuperscript{3*}}

\address{$^1$Department of Electrical and Electronic Engineering, Universidad de Los Andes, Bogot\'a, Colombia}
\address{$^2$Department of Electronics, Universidad de Nari\~{n}o, Pasto, Colombia}
\address{$^3$Department of Biomedical Engineering, Universidad de Los Andes, Bogot\'a, Colombia}
\ead{lf.giraldo404@uniandes.edu.co}
\vspace{10pt}

\begin{abstract}
This paper presents the first study of the long-term impact of climate change on photovoltaic potential in South America. This region has great potential for implementing renewable energy, mainly solar energy solutions, due to its high solar irradiance levels. Based on the Coordinated Regional Downscaling Experiment (CORDEX) for the South American region, we estimate how climate change could affect photovoltaic power potential by the end of the century. The evidence suggests that photovoltaic potential could have a maximum decrease of $15\%$, and a maximum increase of $7\%$, primarily due to changes in solar irradiance of different zones. Furthermore, it is observed that regions with increased temperature also show increased solar irradiance levels, which could, to some extent, compensate for the losses caused by the rise in temperature. Therefore, photovoltaic production in most of the territory will not be negatively affected.
\end{abstract}

%
\vspace{2pc}
\section{Introduction}
Climate change has caused widespread effects on the environment, economies, and communities. Addressing the impact of climate change requires a global effort to reduce greenhouse gas emissions and transition to a more sustainable and resilient future. Renewable energy sources (RES), particularly solar photovoltaic (PV), are key in combating the adverse effects of climate change due to their low operating cost, scalability, zero generation emissions, improved efficiency, and increasingly competitive costs \cite{narvaez2022impact, irena2022world, chu2012opportunities}. However, the efficiency of PV panels depends on meteorological conditions, and climate change could affect PV production \cite{Sampaio2017PhotovoltaicFramework}. Therefore, it is necessary to assess how climate change can affect PV production in the long term \cite{Scherer2016HydropowersFootprint, Hansen2013AssessingNature}. 

The impact of climate change on global PV energy production has been addressed in \cite{wild2015projections, crook2011climate, feron2021climate, HA2023, gernaat2021climate}. These studies used coarse-resolution global climate models (GCM) from the Coupled Model Intercomparison Project \cite{eyring2016overview, yalew2020impacts}. Some works have used regional climate models (RCM), such as the Coordinated Regional Downscaling Experiment (CORDEX) to better approximate a regional scale. Both global and regional climate models are based on representative concentration pathways (RCPs) that estimate different greenhouse gas emission scenarios. The RCP2.6 scenario is optimistic, based on low carbon dioxide emissions and estimates a theoretical limit of a $2^\circ$ increase in global average temperature by the end of the century \cite{van2011rcp2}. The RCP8.5 scenario is pessimistic, assuming high carbon dioxide emissions, resulting in a temperature increase of approximately $4.3^\circ$ relative to pre-industrial temperatures by 2100 \cite{schwalm2020rcp8}. 

Most efforts to understand the impact of climate change on PV energy production have been focused on Europe \cite{Jerez2015TheEurope}, Asia \cite{Zhao2020ImpactsChina}, North America \cite{bartos2015impacts}, Africa \cite{Bichet2019PotentialExperiments, Sawadogo2020ImpactsAfrica}, and Australia \cite{poddar2021estimation}. The PV sector in South America has been growing in recent years, with several countries increasing their solar energy capacity. While Brazil has the largest existing solar infrastructure in the region, Chile has the highest growth rate of solar infrastructure \cite{suri2020global, irena_capacity}. Some studies have attempted to estimate the PV potential in South America \cite{gil2020photovoltaic, gonzalez2021evaluating, hinestroza2021using} and in specific South American regions \cite{narvaez2022impact, Rodriguez-Urrego2018PhotovoltaicProspects, molina2017solar, rios2021selection, de2013grid, cevallos2018spatial}. However, none of these research projects assess the impact of climate change on the PV potential for the entire South American continent. 

This paper evaluates the impact of climate change on PV power potential in South America. We analyze long-term trends in variables affecting PV power potential, such as solar irradiance, air temperature, and wind speed, based on RCP2.6 and RCP8.5 climate scenarios. The study compares a future scenario (2070 -- 2099) with the reference period (1970--1999) using six climate change ensembles from the South America Coordinated Regional Downscaling Experiment (CORDEX) database. The results showed that the PV potential could vary from a decrease of $14.6\%$ to an increase of $6.6\%$, mainly due to changes in solar irradiance according to the RCP8.5 scenario. We identify the regions in South America with the highest solar PV potential and those most affected by climate change. Increases in solar irradiance and temperature are observed in the north of South America, while the south of the continent does not show significant change.

\section{Methods}\label{sec:methods}
To assess the impact of climate change on PV potential in South America, we compare the difference in PV power potential between the reference period (1970-1999) and the estimated PV power potential at the end of the century (2070-2099), following the formulation presented in \cite{narvaez2022impact, Jerez2015TheEurope, Zhao2020ImpactsChina, perez2019climate}. We considered the periods 1970-1999 and 2070-2099 to match with a 30-year life cycle of the PV modules \cite{Jordan2012PhotovoltaicPreprint}. We describe below the climate change models and the PV power potential formulation.

\textbf{Climate change models.} The CORDEX-Coordinated Output for Regional Evaluations (CORE) provides high-resolution regional climate model (RCM) projections \cite{CORDEX}. In CORDEX-CORE, two RCMs were used to downscale four global climate models (GCMs) under two climate scenarios. Table \ref{tab:models_CORDEX} shows the available GCM and RCM models for South America \cite{cordex_core}. 

\begin{table*}
\centering
\caption{Overview of the analyzed CORDEX-CORE experiments. Each experiment has one historical and two scenarios (RCP2.6 and RCP8.5), spanning the periods 1970-2005 and 2070-2099 respectively. The horizontal resolution of all simulations is 0.22º in both latitude and longitude.}
\label{tab:models_CORDEX}
\begin{tabular}{lcc} 
\cmidrule[\heavyrulewidth]{2-3}
 & \multicolumn{2}{c}{\textbf{RCM}} \\ 
\toprule
\textbf{\textbf{Forcing GCM run}} & \multicolumn{1}{l}{GERICS-REMO2015} & \multicolumn{1}{l}{ICTP-RegCM4-7} \\ 
\hline
MOHC-HadGEM2-ES & \usym{1F5F8} & \usym{1F5F8} \\
MPI-M-MPI-ESM-LR & \usym{1F5F8} &  \\
MPI-M-MPI-ESM-MR &  & \usym{1F5F8} \\
NCC-NorESM1-M & \usym{1F5F8} & \usym{1F5F8}
\end{tabular}
\end{table*}

Several studies have evaluated the accuracy of CORDEX-CORE models \cite{giorgi2022cordex, teichmann2021assessing, sawadogo2021current}. These climate change models are based on representative concentration pathways (RCPs) estimating different greenhouse gas emission scenarios. The RCP2.6 scenario is an optimistic case where the global mean temperature increases by a maximum of $2^\circ$ by the end of the century. This scenario could be achieved with a large penetration of renewable energy sources and a forceful reduction of fossil fuels. On the other hand, the RCP8.5 represents the worst-case scenario, where global mean temperature increases by a maximum of $4^\circ$ by the end of the century. This scenario assumes a continuous rise in greenhouse gas emissions, low penetration of renewable energy sources, and intensive use of fossil fuels and coal.

\textbf{PV power potential formulation.} The PV power potential can be expressed as a dimensionless quantity representing the performance of PV cells in relation to ambient conditions and their rated power capacity. Surface-downwelling shortwave radiation (RSDS), surface air temperature (TAS), and surface wind speed (WS) affect the performance of PV cells \cite{Chenni2007ACells}. Higher irradiance leads to an increase in the output current, improving the PV panel's performance. However, a higher PV cell temperature (affected by air temperature, solar irradiance, and wind speed) reduces the output voltage, reducing the efficiency of the PV panel. In this study, we used monocrystalline silicon panels due to their efficiency and widespread use in the industry \cite{Sampaio2017PhotovoltaicFramework}. According to \cite{Jerez2015TheEurope}, the PV power potential ($PV_{pot}$) can be expressed as
\begin{eqnarray}
\label{eq:PVpot}
	PV_{pot} = PR \frac{RSDS}{RSDS_{STC}},
\end{eqnarray}
where RSDS is the downward surface shortwave radiation, STC represents standard test conditions ($RSDS_ {STC}=1000\ W/m^2$), and $PR$ is the performance ratio, which takes into account changes in the efficiency of the PV cells due to changes in their temperature. This ratio is defined as
\begin{eqnarray}
\label{eq:PR}
	PR = 1 + \gamma(T_{cell}-T_{STC}),
\end{eqnarray}
where $\gamma = -0.005^{\circ} C^{-1}$ is a constant that depends on physical parameters of PV cells \cite{Chenni2007ACells}, and $T_{STC} = 25^{\circ}C$. The cell temperature $T_{cell}$ is formulated as
\begin{eqnarray}
\label{eq:Tcell}
	T_{cell} = c_1 + c_2 TAS + c_3 RSDS +c_4 WS,
\end{eqnarray}
with $c_1=4.3^{\circ}C$, $c_2=0.943$, and $c_3=0.028^{\circ}Cm^2/W, c_4=-1.528^{\circ}Cs/m$ being constants that depend on physical parameters of the PV cells \cite{Chenni2007ACells}. Moreover, $TAS$ is the surface air temperature, and $WS$ is the surface wind speed. According to Equations (\ref{eq:PVpot}) and (\ref{eq:PR}), if environmental conditions are the same as standard conditions, the PV power output reaches its nominal value ($PV_{pot}=1$). If the air temperature is higher than the nominal temperature ($T_{cell}>25^{\circ}C$) or the solar irradiance is lower than the nominal one ($RSDS <1000 W/m^2$), the PV power output will be less than the rated power of the module ($PV_{pot}<1$). On the other hand, if the temperature is lower than the nominal temperature ($T_{cell} <25$ ), or if the irradiance is higher than the nominal irradiance ($RSDS> 1000$), the nominal generation of the PV panel will improve ($PV_{pot}>1$).

Using Equations (\ref{eq:PVpot}) - (\ref{eq:Tcell}) $PV_{pot}$ can be expressed as:
\begin{eqnarray}
\label{eq:PVpot_group}
	PVpot = RSDS(a + b RSDS + c TAS  + d WS),
\end{eqnarray}
where 
\begin{eqnarray*}
a&=\frac{1+\gamma(c_1-T_{STC})}{RSDS_{STC}},\\
b&=\frac{\gamma c_3}{RSDS_{STC}},\\
c&=\frac{\gamma c_2}{RSDS_{STC}}, \\
d&= \frac{\gamma c_4}{RSDS_{STC}}.
\end{eqnarray*}
From Equation \ref{eq:PVpot_group}, changes in $PV_{pot}$, denoted as $\Delta PV_{pot}$, can be calculated as:
\begin{eqnarray}
\label{eq:DeltaPVpot}
    \Delta PV_{pot} = 
    &
    \Delta RSDS \left(a + b\Delta RSDS + 2bRSDS + cTAS + dWS \right) \\
    &
    +cRSDS \cdot \Delta TAS \nonumber \\
    &
    +dRSDS \cdot \Delta WS \nonumber\\
    &
    +c \Delta RSDS \cdot \Delta TAS \nonumber\\
    &
    +d \Delta RSDS \cdot \Delta WS \nonumber
\end{eqnarray}

The changes in $PV_{pot}$ due to the individual contributions of $\Delta RSDS$, $\Delta TAS$, and $\Delta WS$ can be calculated by Equation (\ref{eq:DeltaPVpot}). For example, the change in $PV_{pot}$ caused by variations in $TAS$ is given by setting $\Delta RSDS = \Delta WS = 0$ and keeping the remaining variables ($RSDS$, $WS$) constant at their annual mean during the reference period (1970-1999). We follow the same methodology as the one presented in \cite{narvaez2022impact, Jerez2015TheEurope, perez2019climate}. It is worth noting that it is not possible to completely isolate the contribution of each variable due to the cross-products in the last two terms of Equation (\ref{eq:DeltaPVpot}).

\section{Results}\label{sec:results}
\begin{figure*}
    \centering
    \includegraphics[width=1\textwidth]{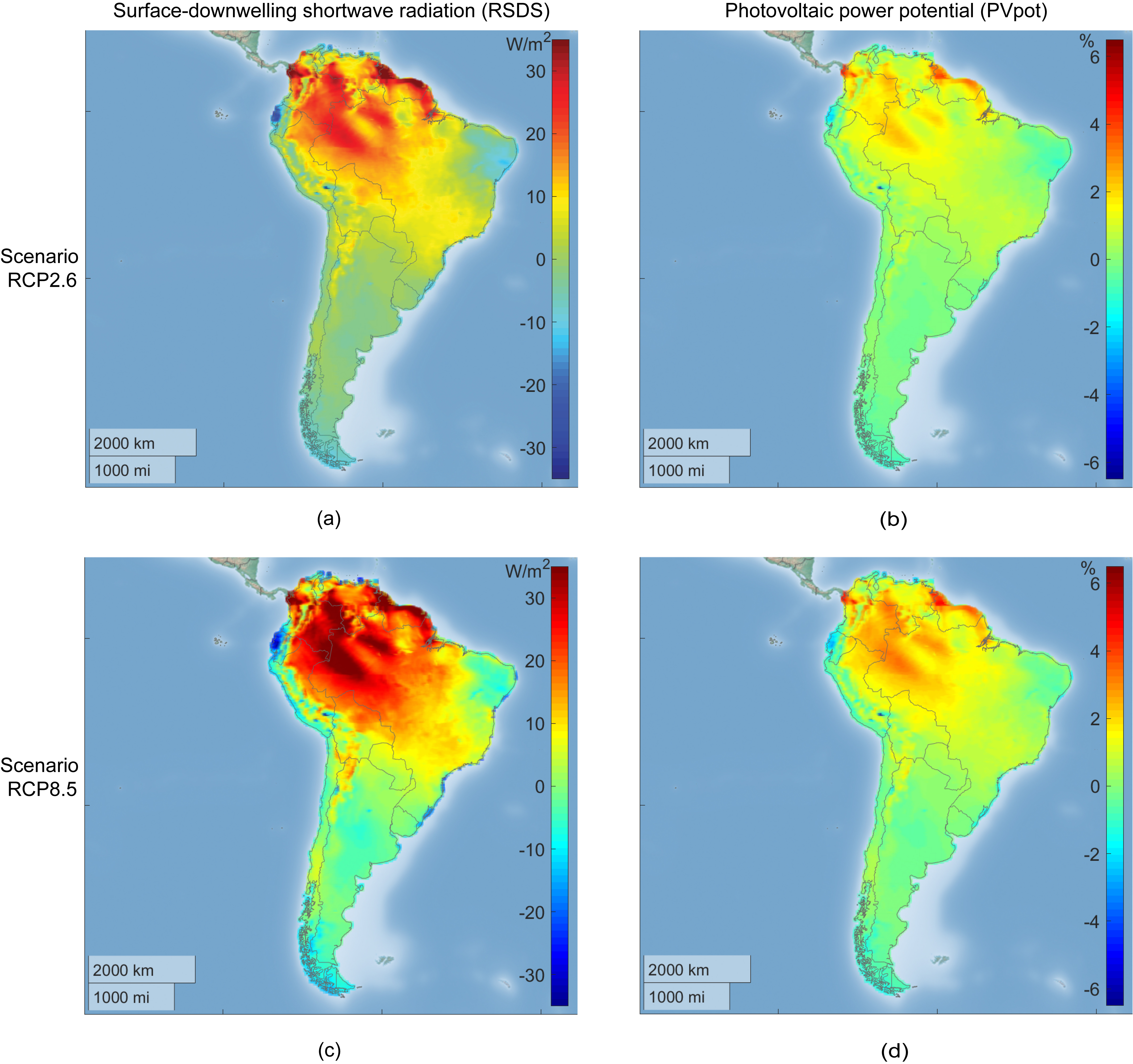}
    \caption{{\bf Solar irradiance change analysis.} The left column compares the change in solar irradiance, in Watts per square meter, between the end of the century (2070-2099) and the reference period (1970-1999). The right column represents the respective change in PV power potential in percentage. In particular, (a) changes in RSDS with RCP2.6, (b) $PV_{pot}$ with RCP2.6, (c) changes in RSDS with RCP8.5, and (d) $PV_{pot}$ with RCP8.5. The plots were generated in Matlab.}
    \label{fig_RSDS}
\end{figure*}

\textbf{Change in solar irradiance and its influence on PV power potential.} Surface-downwelling shortwave radiation (RSDS) represents the amount of energy that reaches the Earth's surface from the sun. Fig. \ref{fig_RSDS}(a) and Fig. \ref{fig_RSDS}(b) show, according to the RCP2.6 scenario, changes in RSDS between the historical reference period (1970-1999) and the projected end of the century period (2070-2099), and the effect on PV power potential in South America, respectively. The southern part of the continent shows no significant changes. In contrast, the northern part experiences a general increase of about $30W/m^2$ in most areas, with a maximum increase of 64$W/m^2$ in northern Colombia. However, some regions, particularly western Ecuador, experience a decrease in solar irradiance of around 30$W/m^2$. The greatest decrease is observed in southern Peru, with a value of 112$W/m^2$. This decrease is much larger than the next largest decrease (Bolivia with $42W/m^2$) and the average decrease for each country ($36W/m^2$). Therefore, combining the accuracy of historical data with current in-situ measurements as well as improving this database with site-adaptation techniques as proposed \cite{Narvaez2021MachineForecasting, gaviria2022machine}, would be of great value. These solar irradiance variations may cause a maximum decrease in the PV potential of $12\%$ in southern Peru, western Ecuador, and some coastal regions of northeastern Brazil (shown in blue on the map). However, an increase of up to $6\%$ is observed in the northern region (shown in red on the map), and on average, the continent shows an increase of $8W/m^2$, resulting in an expected increase of $0.6\%$ on $PV_{pot}$ across the continent.

The effects on RSDS and PV power potential in South America, according to the RCP8.5 scenario, are shown in Fig. \ref{fig_RSDS}(c) and Fig. \ref{fig_RSDS}(d), respectively. Solar irradiance generally increases by over $30W/m^2$ in the northern region of the continent, with a maximum increase of $74W/m^2$ in northern Colombia. Some coastal areas, especially in Ecuador, present a reduction in solar irradiance of around $30W/m^2$. As in the previous scenario, the greatest decrease is observed in southern Peru, with a reduction of $130W/m^2$. On average, the continent shows an increase of $11W/m^2$ in solar irradiance. According to this scenario, the variations in solar irradiance could cause a maximum decrease of $15\%$ on PV power potential (indicated by blue regions on the map), while an increase of up to $7\%$ is observed in the northern region (indicated by red regions on the map). An average increase of $1\%$ on $PV_{pot}$ across the continent is expected due to an average increase of $11.3W/m^2$ in solar irradiance.

\begin{figure*}
    \centering
    \includegraphics[width=1\textwidth]{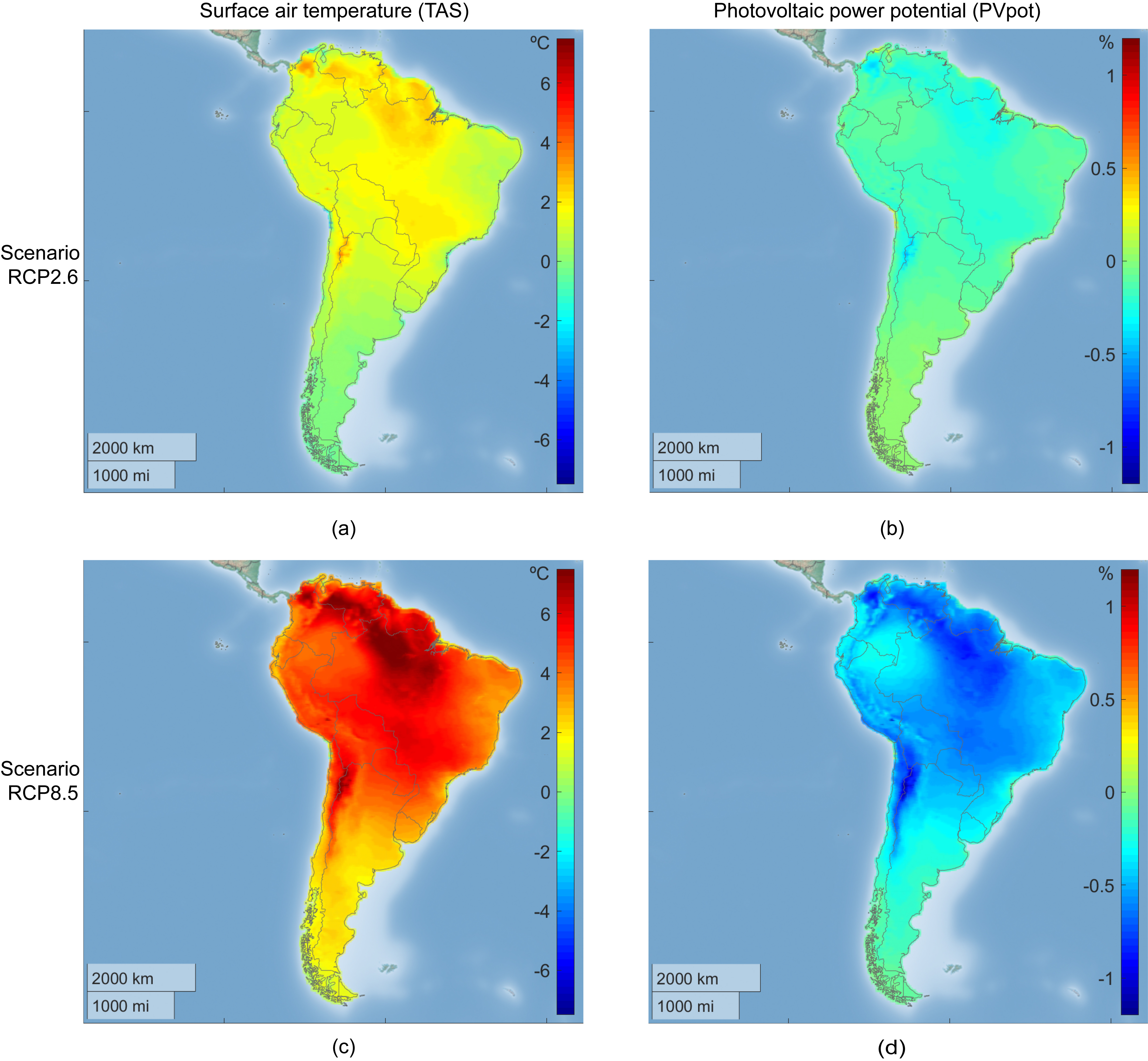}
    \caption{{\bf Temperature change analysis.} The left column compares the change in temperature (in Celsius) between the end of the century (2070-2099) and the reference period (1970-1999). The right column represents the respective change in PV power potential in percentage. In particular, (a) Changes in TAS with RCP2.6, (b) $PV_{pot}$ with RCP2.6, (c) TAS with RCP8.5, and (d) $PV_{pot}$ with RCP8.5. The plots were generated in Matlab.}
    \label{fig_temperature}
\end{figure*}

\textbf{Change in temperature and its influence on PV power potential.} The effects on surface air temperature (TAS) and PV power potential in South America, according to the RCP2.6 scenario, are shown in Fig. \ref{fig_temperature}(a) and Fig. \ref{fig_temperature}(b), respectively. There is a general increase in the air temperature, especially on the northern part of the continent, with a rise of approximately $2^\circ$ and a maximum increase of $7^\circ$ on the border between Bolivia, Chile, and Argentina. In the southern part of the continent, the temperature increase is around $1^\circ$. However, the results report a decrease of $4^\circ$ in a small area on the northern coast of Chile. The average temperature rise for the entire continent is $1.4^\circ$. These changes can affect PV power potential, resulting in a $1\%$ decrease in PV cell performance in warmer locations, and an average reduction of $0.15\%$ across the continent.

Fig. \ref{fig_temperature}(c) and Fig. \ref{fig_temperature}(d) present the effects of changes in air temperature and PV power potential based on the RCP8.5 scenario. In this scenario, the changes are more notable than in the case of RCP2.6, with a maximum temperature increase of $11.3^\circ$ and a maximum temperature decrease of $1.9^\circ$. The region most affected by these changes is the northwest of South America, where a decrease in PV power potential of up to $1.8\%$ is observed. On average, the temperature across the entire continent rises by $4.6^\circ$, resulting in an average decrease of $0.5\%$ in PV power potential.

\begin{figure*}
    \centering
    \includegraphics[width=1\textwidth]{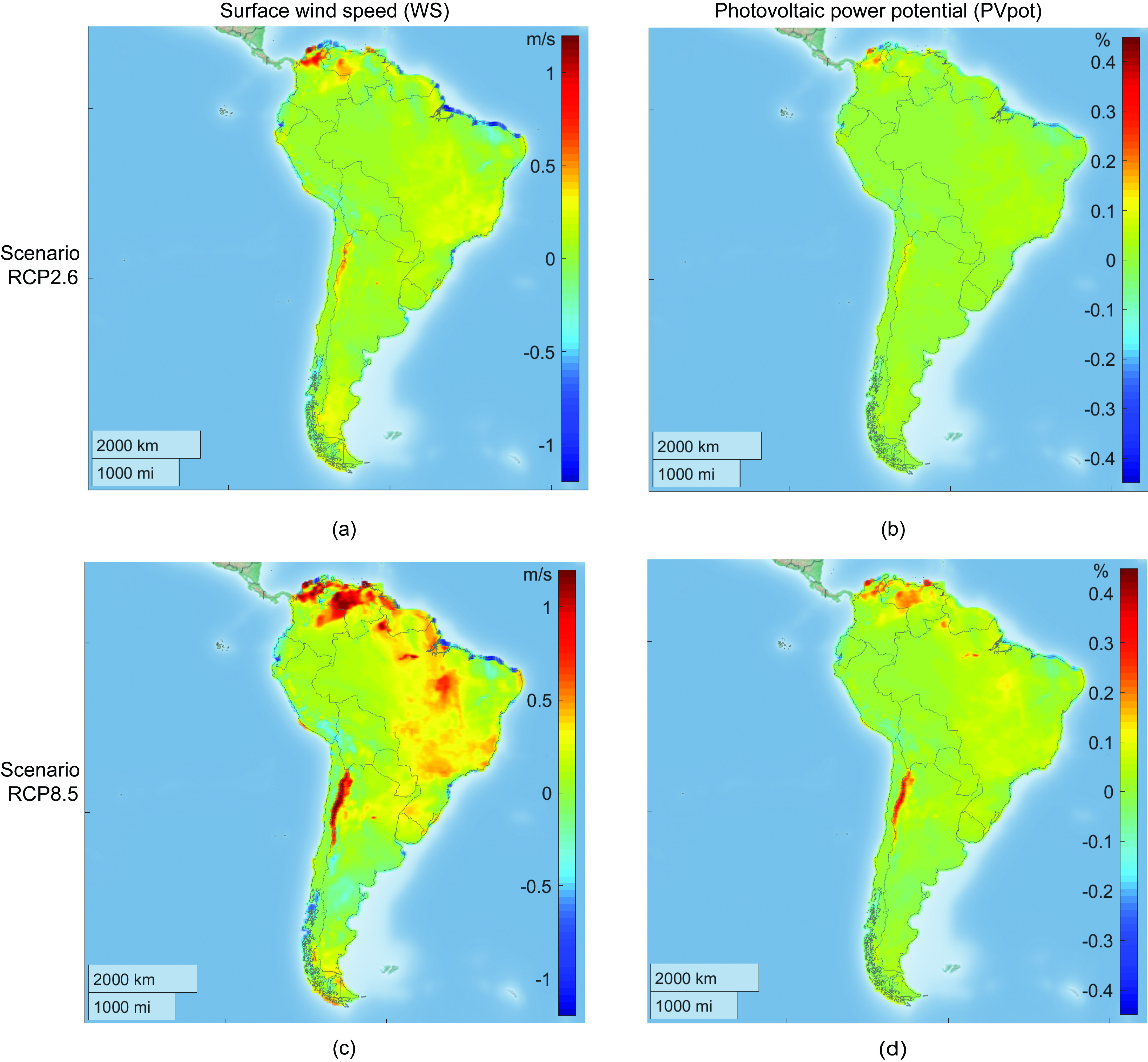}
    \caption{{\bf Wind change analysis.} The left column compares the change in wind speed (in meters per second) between the end of the century (2070-2099) and the reference period (1970-1999). The right column represents the respective change in PV power potential in percentage. In particular, (a) Changes in WS with RCP2.6, (b) $PV_{pot}$ with RCP2.6, (c) WS with RCP8.5, and (d) $PV_{pot}$ with RCP8.5. The plots were generated in Matlab.}
    \label{fig_wind}
\end{figure*}

\textbf{Change in wind speed and its influence on PV power potential.} Fig. \ref{fig_wind}(a) and Fig. \ref{fig_wind}(b) show, according to the RCP2.6 scenario, the projected changes in wind speed and their effects on PV potential, respectively. In general, wind speed changes range from $-1.5m/s$ to $1.3m/s$. These changes have a practically negligible impact on PV power potential, with variations between $-0.28\%$ and $0.26\%$.

For the RCP8.5 scenario, the results are presented in Fig. \ref{fig_wind}(c) and Fig. \ref{fig_wind}(d). In this scenario, changes in wind speed range from $-1.3m/s$ to $1.95m/s$, which may affect the PV power potential between $-0.28\%$ and $0.43\%$. These results demonstrate the slight impact of wind speed on the PV power potential and are in accordance with other results presented in the literature \cite{Jerez2015TheEurope, Sawadogo2020ImpactsAfrica, Bichet2019PotentialExperiments, carreno2020potential,perez2019climate}.

\textbf{Changes in $PV_{pot}$ by country.} The following is an analysis of the impact of climate change in the most populated South American countries based on the RCP8.5 scenario.

\textbf{Argentina.} The results suggest that the northern part of the country will have an overall increase in air temperature between $3^\circ$ and $4^\circ$, with the most significant increase near the border with Chile and Bolivia ($9^\circ$). Argentina is projected to have an average temperature increase of $3^\circ$ throughout the country. These results are consistent with those presented in \cite{barros2015climate}. This increase in air temperature will have a negative impact on the country's PV potential, resulting in a general reduction of $0.3\%$ and a maximum reduction of $1.4\%$ in the warmer regions. Changes in solar irradiance vary between $-23W/m^2$ and $25W/m^2$, leading to changes in PV potential of $-2.6\%$ and $2.5\%$, respectively.

\textbf{Bolivia.} The overall temperature increase in the country would be $5.1^\circ$ with a maximum increase of $11.3^\circ$. These changes could result in reductions in PV potential by $0.6\%$ and $1.8\%$, respectively. However, a general increase of $10.7W/m^2$ in solar irradiance is expected. These air temperature and solar irradiance projections are aligned with those presented in \cite{seiler2013likely}. This increase in solar irradiance would result in a $1\%$ increase in PV potential, compensating to some extent for losses due to temperature increase.

\textbf{Brazil.} The minimum temperature increase in the coastal zone of Brazil would be $0.4^\circ$, while the northern zone could experience increases of up to $8.2^\circ$. The average temperature increase for the entire country is $5.1^\circ$, resulting in a $0.5\%$ to $1\%$ decrease in PV potential. Changes in solar irradiance range from $-29W/m^2$ on the coast and $42W/m^2$ in the northwest, resulting in a decrease in PV potential of up to $2.5\%$, and an increase of up to $3.5\%$. These climate change projections for Brazil could also impact other energy sectors, such as hydropower generation and liquid biofuel production, as shown in \cite{de2009vulnerability}.

\textbf{Chile.} Southern Chile shows a decrease in temperature of up to $2^\circ$, as well as a decrease in solar irradiance of up to $22W/m^2$. These changes in solar irradiance could cause a decrease in the PV potential of up to $2.6\%$. However, the decrease in temperature could increase the PV potential by up to $0.2\%$. On the other hand, northern Chile exhibits an increase in both temperature (with a maximum of $8.5^\circ$) and solar irradiance (with a maximum of $17W/m^2$), causing the PV potential to decrease by $1.3\%$, and to increase by $1.7\%$, respectively.

\textbf{Colombia.} Solar irradiance levels show a generalized decrease, especially in the southwest, with a maximum decrease of $27W/m^2$. The southwest of Colombia presents the smallest increase in temperature; however, this increase is considerable with values of around $3^\circ$. These changes could cause decreases in PV potential varying from $0.5\%$ to $2.4\%$. These results are consistent with those presented in \cite{narvaez2022impact}. The southeast and northwest would experience solar irradiance increases (with a maximum of $73.5W/m^2$), resulting in an increase in PV potential of up to $6.6\%$. However, these regions could experience a decrease in PV potential of up to $1\%$ due to temperature increase (with a maximum of $7.4^\circ$). On average, the entire Colombian territory would experience an increase in solar irradiance of about $25W/m^2$, resulting in a $2\%$ increase in PV potential. The temperature will increase by $4.7^\circ$, leading to a decrease in PV potential of $0.5\%$.

\textbf{Ecuador.} The coast would present decreases in solar irradiance levels (maximum $38W/m^2$), which would cause drops in PV potential of up to $3.1\%$. This zone shows temperature increases ranging from $2^\circ$ to $3^\circ$, affecting the PV potential between $0.2\%$ and $0.4\%$. On the other hand, the eastern side of the country would exhibit increases in solar irradiance ($31W/m^2$), but also higher temperature increases ($4.4^\circ$). These changes would have a positive effect on PV potential by up to $2.7\%$ due to irradiance increases, but also negatively affect it by up to $0.6\%$ due to higher temperatures. On average, solar irradiance would decrease by $1.6W/m^2$, and temperature would increase by $3.6^\circ$. These changes would decrease the PV potential by $0.2\%$ and $0.4\%$, respectively.

\textbf{Paraguay.} Paraguay could experience temperature increases of between $4^\circ$ and $5.6^\circ$, which would cause reductions in PV potential of about $0.5\%$. These changes are similar to those presented in \cite{silvero2020energy}. Solar irradiance would not undergo significant changes, with variations between $-5.7W/m^2$ and $9.1W/m^2$, affecting the PV potential between $-0.5\%$ and $0.8\%$.

\textbf{Peru.} The results show that a small region in southern Peru would drastically decrease solar irradiance levels by $-130W/m^2$, leading to a decrease in PV potential of up to $14.6\%$. It would be worthwhile to examine the accuracy of the irradiance data for this particular region, as these changes differ significantly from those observed in the rest of the country and the continent. The maximum increase in solar irradiance is up to $38W/m^2$, which can boost the PV potential by up to $3\%$. This increase is most notable in the northeast of the country. The average temperature across the entire Peruvian territory presents an increase of $4.3^\circ$, leading to a decrease in PV potential by up to $1.2\%$. However, the models show a decrease in temperature in the northwest of up to $1.4^\circ$, which could have a positive effect on PV potential by up to $0.2\%$.

\textbf{Uruguay.} Uruguay is the country with the least drastic changes. The temperature would vary between $1.4^\circ$ and $3.7^\circ$, while irradiance would vary between $-15W/m^2$ and $1.2W/m^2$. These variations affect the PV potential, which can decrease by up to $-1.4\%$ or increase by up to $0.1\%$.

\textbf{Venezuela.} Solar irradiance would vary from $-34W/m^2$ to $58W/m^2$. Increases in irradiance are mainly observed in the east and west, while decreases are observed in the northern part of the country. These changes in irradiance would affect the PV potential by $-3\%$ to $4.7\%$. The temperature across the country would increase on average by $5.7^\circ$, with a maximum of $7.7^\circ$ in the center and west, affecting the PV potential by up to $0.8\%$.

\section{Discussion}\label{sec:discussion}
South America has great potential for renewable energy solutions, mainly due to high levels of solar irradiance. It is important to note that countries such as Colombia, Venezuela, Ecuador, Peru, Brazil, Bolivia, and Uruguay have tropical or subtropical climates. This condition causes temperature and irradiance values to remain relatively constant throughout the year, making it ideal for the use of solar energy through fixed-tilt PV systems. In particular, Chile, Bolivia, and Argentina are among the top ten countries in the world with the highest levels of solar irradiance \cite{meisen2009potencial}. Despite the great potential of solar and wind energy solutions, South America's energy supply depends primarily on hydropower. This over-reliance on hydropower has caused supply problems during periods of drought, leading to power failures. However, solar and wind power are taking a significant share of the electricity portfolio, with exponential growth in the last decade. This growth is expected to continue in the medium and long term due to the policies established by most South American countries \cite{barbosa2017hydro, gil2020photovoltaic, da2021power}.

According to \cite{da2021power}, solar energy is expected to be the primary renewable energy source in South America in the upcoming decades. Therefore, it is crucial to evaluate both, the photovoltaic power potential and its susceptibility to climate change. The following is a comprehensive analysis of the potential impact of climate change on photovoltaic energy throughout the entire South American region.

According to our results, the variable that would have the greatest effect on PV potential in South America at the end of the century is solar irradiance. It would generate a maximum decrease of $15\%$ and a maximum increase of $7\%$ in $PV_{pot}$ based on the RCP8.5 scenario, and $12\%$ and $6\%$ based on the RCP2.6 scenario. These changes in solar irradiance would positively affect $PV_{pot}$ in the northeastern region of South America, especially in southeastern Colombia, southwestern Venezuela, eastern Ecuador, northeastern Peru, and western Brazil. On the other hand, some regions that would have a decrease in $PV_{pot}$ are southwestern Colombia, western Ecuador, southern Peru, and southeastern Brazil. However, this reduction would occur only in small regions of the continent. Therefore, this decrease in solar irradiance can be considered a small threat to the future of PV solutions in South America.  

The air temperature is the second most important variable affecting PV potential. The differences between the RCP2.6 and RCP8.5 scenarios are remarkable. While the RCP2.6 scenario estimates a less dramatic temperature increase, the RCP8.5 scenario predicts a more intense one. According to these air temperature changes, $PV_{pot}$ would decrease by a maximum of $1\%$ to $1.8\%$ under RCP2.6 and RCP8.5 scenarios, respectively. 

The northern region of the continent would experience the most critical increase in air temperature, especially in northeastern Colombia, western Venezuela, northwestern Brazil, northern and southern Bolivia, northern Paraguay, northeastern Chile, and northwestern Argentina. Although temperature changes are expected to decrease the $PV_{pot}$ in these areas, this decrease would be less than $2\%$. In addition, an increase in solar irradiance is expected in some of these regions, which could offset the losses caused by the increase in air temperature. Therefore, the photovoltaic potential for this region is still positive. However, it is important to note that an increase above $4^\circ$ would exceed the upper limit of $2^\circ$ by the end of the century \cite{Peters2013TheC, huntingford2012link}, causing possibly irreversible changes to nature. Consequently, a joint effort among all nations is necessary to achieve a transition to clean energy sources and thus reduce global warming, as suggested in the latest Intergovernmental Panel on Climate Change (IPCC) report \cite{IPCC}. South America can be a leader in this transition for a $100\%$ renewable energy supply as proposed in \cite{barbosa2017hydro}.

It is worth noting that areas of projected air temperature increases (which cause decreases in photovoltaic potential) are somewhat compensated for by increases in solar irradiance.

The projected changes in wind speed for South America do not represent a significant threat. Other papers report similar results regarding the influence of wind speed on PV potential \cite{Jerez2015TheEurope, Sawadogo2020ImpactsAfrica, Bichet2019PotentialExperiments, carreno2020potential,perez2019climate}. However, it would be worth examining whether changes in wind speed affect other renewable energy sources such as wind energy. These changes are most notable in the RCP8.5 scenario. Northern and eastern Colombia, western Venezuela, northern and central Brazil, and the northern part of Chile on the Argentinean border could see increases of up to $1m/s$.

While climate models have been proven to be reliable tools for projecting many aspects of climate change, modeling the impacts water bodies have on climate models remains an active area of research \cite{kundzewicz2018uncertainty, nobrega2011uncertainty}. The climate change models used in this study have shown significant changes in solar irradiance and temperature in places where there are rivers, lakes, and coastlines. Therefore, it is essential to validate the effectiveness of climate change models in areas with bodies of water \cite{fantini2018assessment, luhunga2016evaluation}. Efforts should be made to improve the representation of water bodies in climate models, such as incorporating more detailed observations and advanced modeling techniques. Additionally, this study did not consider the orientation and tilt of the PV panels, which is also a significant factor affecting their performance. Moreover, when deciding on the best locations for implementing PV systems, land-use restrictions, such as nature reserves, should be taken into account.

\section*{Acknowledgment}
This work was supported by Fundaci\'on CEIBA through the program B\'ecate Nari\~{n}o, and by Centro de Desarrollo Sostenible para Am\'erica Latina through convocatoria para financiaci\'on de proyectos de investigaci\'on relacionados con el alcance de los objetivos de desarrollo sostenible. All authors acknowledge the financial support provided by the Vice Presidency for Research and Creation publication fund at the Universidad de Los Andes, Colombia. 

\end{document}